\documentclass[a4paper]{article}
\usepackage{setspace}   

\usepackage[english]{babel}
\usepackage[utf8x]{inputenc}
\usepackage[T1]{fontenc}
\usepackage[a4paper,top=3cm,bottom=2cm,left=3cm,right=3cm,marginparwidth=1.75cm]{geometry}

\usepackage{amsmath}
\usepackage{graphicx}
\usepackage[colorinlistoftodos]{todonotes}
\usepackage[colorlinks=true, allcolors=blue]{hyperref}

\title{A Neuronal Noise Critique of Integrated Information Theory}
\author{Refath Bari}

\begin{document}
\maketitle

\section{Abstract}
Integrated Information Theory (IIT) is an audacious attempt to pin down the abstract, phenomenological experiences of consciousness into a rigorous, mathematical framework. We show that IIT's stance in regards to neuronal noise is inconsistent with experimental data demonstrating that neuronal noise in the brain plays a critical role in learning, visual recognition, and even categorical representation. IIT predicts that entropy due to noise will reduce the information integration of a physical system, which is inconsistent with experimental data demonstrating that decision-related noise is a necessary condition for learning and visual recognition tasks. IIT must therefore be reformulated to accommodate experimental evidence showing both the successes and failures of noise.

\section{Introduction}
The very definition of consciousness is subject to much debate, and includes everything from consciousness is the way information feels when it's being processed \cite{53} to consciousness does not exist \cite{54} -- it is merely a term concocted to describe a set of processes we do not yet fully understand, akin to elán vital in the early 20th century \cite{55}. As the definition itself is hotly contested, it comes as no surprise that responses to consciousness spread far and wide, from those arguing that consciousness is the only real thing and all else is manifest from it, to consciousness is an imagined entity constructed to explain an unknown functional process in the brain. One group, panpsychists, claim consciousness is universal, ingrained in everything from electrons to tables to humans \cite{28}; reductionists argue consciousness is equal to the sum of its parts, explainable solely by the machinery of the human brain \cite{29}; dualists profess the world is no more than a manifestation of the mind, and that the mind and brain are two separate entities \cite{30}; and eliminativists proclaim consciousness doesn't exist at all, only physical systems and their form/function \cite{27}.

\space
\bigskip

The subject of consciousness has been too taboo for neuroscience, until just recently -- after all, how can one mathematically formulate an entity so abstract and intangible? It's no surprise, then, that consciousness is absent in everything from the periodic table to Relavity/Quantum Mechanics to the entire human genome \cite{38}. It seems — in the pursuit of objectiveness — we've managed to neglect the very essence of who we are: our own subjective experiences, and the consciousness that makes it possible. But in the late 20th century, with the advent of electroencephalography (EEG), magnetic resonance imaging (MRI), and functional MRI (fMRI), for the first time, neuroscientists could expose a subject to a stimuli and watch the brain respond in real-time. This opened the door for all sorts of experiments involving the brain's conscious precepts, and finally enabled a neurobiological approach to cracking the problem of consciousness \cite{39}\cite{40}.

\section{Integrated Information Theory (IIT)}
In 2004, Gulio Tononi, a neuroscientist at the University of Wisconsin-Madison, proposed a full-fledged theory of consciousness: Integrated Information Theory (IIT) \cite{18}. It was an audacious attempt to pin down the abstract, phenomenological experiences of consciousness into a rigorous, mathematical framework. Tononi devised five axioms universal to all subjective experiences: consciousness exists, it has structure, it is unique, it is integrated, and it is exclusive. The first axiom was founded on Descartes’ cogito ergo sum: “I think, therefore I am” \cite{42}. In other words, consciousness is a fundamental ingredient to all entities, almost in a panpsychic sense. The second axiom states that consciousness is the product of components, such as color, size, and shape. The third axiom states that every experience is unique, in the sense that there are countless aspects of any given experience that differentiates it from another one. The fourth axiom states that consciousness is responsible for the synthesis of disparate forms of information, including electromagnetic waves of light, pressure waves of sound, and diffusion of smell. The final axiom is responsible for exclusion: the principle that we do not simultaneously experience multiple conscious experiences -- our consciousness suppresses alternative interpretations of the physical world (i.e., a colorless world, a soundless world, etc.) so that we have one coherent, unified physical reality. IIT 3.0, an updated theory of Information Integration, has proposed eight axioms instead of five, but the central tenets of information integration and exclusion remain the same \cite{17}.

\space
\bigskip

Since 2004, IIT has become the leading theory of consciousness, but remains a minority position among neuroscientists. The key takeaway from IIT is that consciousness is the product of (1) information integration and (2) causal structures in the brain. As explained by the binding problem, the brain consumes disparate inputs and binds them together into a coherent reality. IIT has an important implication in that it states that information can only be integrated by causal structures — i.e., systems where components can have a cause-effect on each other. As a result, feed-forward networks such as artificial feed-foward neural networks would not integrate information, according to IIT, but networks involving feedback loops do integrate. IIT makes an attempt to quantify every aspect of the conscious experience, proposing a single quantity, $\phi$, a quantity which is postulated to have an identity relationship with consciousness. IIT thus implies any physical system which involves a high amount of information integration has a high $\phi$, and thus a high level of consciousness, such as an awake human. A physical system with a low amount of information integration would have a low $\phi$, and thus a low level of consciousness, as in the case of a worm. IIT has made a few predictions -- some of which are clinically testable, and others which remain infeasible (i.e., calculating $\phi$ for any sufficiently large physical system).

\section{Predictions, Successes, and Failures of IIT}
IIT also makes several counter-intuitive predictions about consciousness, such as that a sufficiently integrated logic gate has a greater level of consciousness than humans \cite{45}. This has caused much controversy in regards to the falsibiliality of IIT, as it excludes functionalism as a means of determining whether a system is consciousness, and instead relies solely on $\phi$. IIT leaves much to be desired -- $\phi$ is notoriously difficult to calculate, as it grows super-exponentially with the number of connections in a system. In addition, IIT has been criticized for failing to address basic questions of consciousness \cite{15}\cite{16}\cite{44}, including Chalmers’ hard problem of consciousness. In this paper, we will examine IIT’s stance on neuronal noise and contrast it to recent research demonstrating the benefits of noise for information integration and address the counterargument regarding the neuronal noise hypothesis of cognitive aging.

\space
\bigskip

Computationally-efficient reformulations of IIT’s $\phi$ quantity have been successfully used to distinguish several levels of consciousness in patients, from minimal (i.e., vegetative patients) to normal levels. Furthermore, sleep phenomenologically seems to involve a lower level of consciousness than wakefulness, a phenomena which IIT explains by stating that the brain’s connections are less integrated during sleep -- a prediction which has been empirically corroborated \cite{52}. IIT has also been vindicated in the brains of vegetative patients can interpret and process language, who IIT predicts would have a greater level of consciousness than vegetative individuals who do not respond to external queries \cite{52}.

\space
\bigskip

However, the basic foundations of IIT have been criticized for a multitude of reasons: (1) that the axiom of information integration may be a "necessary, but insufficient" condition for consciousness and (2) Tononi defines information integration in terms of information exclusion without justifying the latter \cite{15}. Furthermore, the principle of information integration may seem self-evident to Tononi, but is far from required for complex systems which consume and process data in such a way that the future state of the system is dynamically dependent on the past state -- indeed, Scott Aaronson provides counterexamples such as linear-size super-concentrators for LDPC codes \cite{45}. Furthermore, Aaronson argued that $\phi$ could potentially have a larger value for an arbitrarily complex logic gate than a human, to which Tononi responded by stating that we must not trust our intuitions when it comes to consciousness. Instead, if IIT predicts that a logic gate has greater consciousness than a human, who are we to say otherwise? Others have criticized the trivial predictions of IIT, as the same prediction are made by a trivial theory such as Circular Coordinated Message Theory (CCMT), which can produce the same explanation for why the cerebellum contains most of the brain's neurons, but accounts for practically none of our conscious experience \cite{15}. 

\space
\bigskip

IIT's defining quantity, $\phi$, which is claimed to be equivalent to the degree of consciousness of a physical system, is notoriously difficult to calculate. Even for a simple physical system such as a worm, with 302 synapses, it would take $10^9$ years to compute $\phi$. Tegmark et al. demonstrated that $\phi$ grows super-exponentially with the number of possible divisions of a system \cite{43}. Many empirical studies use alternative, more computationally-efficient definitions of $\phi$ to test IIT's predictions. However, different definitions of $\phi$ can lead to significantly varying levels of consciousness for any given system. However, proponents of IIT maintain that the theory is hypothetically testable, although not feasible at the time.

\section{A Novel Critique of IIT}
To understand the inconsistency between IIT's stance on neuronal noise and contemporary research on noise, we must first understand the nature and origins of neuronal noise.

\space
\bigskip

Brain function has been demonstrated to significantly vary within the same subject, even when the subject is presented with the same stimuli \cite{5}. This variability is often interpreted as "neuronal noise," although McDonnell and Ward have proposed "stochastic facilitation" to describe such noise when beneficial to cognitive functioning \cite{25}. Neuronal noise can arise from many factors, such as synaptic delay, exhibition-inhibition cycles, synaptic connectivity, rate of cell firing, and many other factors (motor noise, thermal noise, etc.) \cite{1}\cite{3}\cite{8}. For many decades, noise was no more than a nuisance which neuroscientists would minimize by averaging trials of a subject's responses to the same stimuli \cite{9}. In addition, many biologists presumed noise originated from instrumentation, but He et. al. demonstrated that even after controlling for instrument noise, a significant residual variability remained -- neuronal noise \cite{22}. This noise is thus an artifact of the brain itself, and could provide hints as to how the brain generates conscious precepts.  

\space
\bigskip

When measured with an EEG, brain activity produces several types of brain oscillations at different frequencies. Each of these oscillations has its own function in the brain \cite{2}. For instance, the first discovered brain oscillation was the occipital alpha wave, the dominant brain signal when a subject's eyes are closed, and reduced when eyes are open, or the subject is drowsy. In contrast, there are aperiodic signals in the brain, as characterized by the $\frac{1}{f}$ noise in the power spectra of an EEG. For decades, noise was no more than a distraction from dominant brain oscillations, and was actively removed from analyses. This noise was originally assumed to originate from instrumentation, but He et. al. demonstrated that instrumentation accounts for only a small amount of the noise \cite{22}. The rest, therefore, originates from the brain. Voytek et. al. has demonstrated that Aperiodic Noise is dominant in the infant brain, and rapidly changes in the first seven months of infancy, suggesting that Neuronal Noise plays a significant role in cognitive development and aging \cite{20}. In addition, aperiodic brain activity may be relevant to task performance, as demonstrated by Grossman et. al., as well as the state of arousal, as demonstrated by Lender et. al \cite{21}.

\space
\bigskip

Recent studies have demonstrated that neuronal noise is no distraction — indeed, it may be critical to cognitive functioning. Engel et. al. showed that noise is critical to neuronal categorical representation. In two simulated networks that imitate perception and category formation in the brain, Engel added decision-related noise to one and no noise to the other. The latter model failed to create categories and learn, unlike the model equipped with noise, thus demonstrating that where there is no noise, there is no learning \cite{46}. Newsome et. al. previously showed in a study of mice recognizing whether a series of dots are moving to the right or left that the mice must make a "random" first guess, based on decision-related noise, to learn which way the dots are moving \cite{47}. In other words, without a random guess, the brain will never learn how to improve — where there is no mistake, there is no learning. This is quite similar to Artificial Neural Networks, whose weights and biases are randomized at initialization, so that the network makes a few random decisions, and then improves based on stochastic gradient descent. \cite{48}

\space
\bigskip

The Bayesian Coding hypothesis posits that the human brain makes decisions and computations based on a bayesian probability scheme. For instance, imagine the sensation of a drum. If the sound of the drum and the location of the drum are in the same position, there will be a large overlap in the likelihood functions of the visual and auditory cues. However, if one hears the sound of a drum from the left but sees the actual drums on the right of the peripheral vision, there will be less overlap in the likelihood functions, as they provide conflicting cues. Pouget and Knill state that if both the visual and auditory likelihood functions are gaussian distributions, the most likely direction of the drum sound would be given by a weighted sum of the visual and auditory cues, $\mu_{v,a}=w_v\mu_v+w_a\mu_a$. In essence, the brain is like a college admissions committee, constantly making decisions based on a weighted sum of the applicant's appearance, sound, smell, taste, and touch. Ma et. al. has demonstrated that the noise in the brain simplifies complex Bayesian computations into much simpler linear combinations of populations of neural activity \cite{49}\cite{4}. As demonstrated by Pouget, Neuronal Noise fits well with Poisson distributions, which is precisely the type of distribution under which Bayesian computations are conducted most efficiently. \cite{50}

\space
\bigskip

Artificial Neural Networks (ANNs) have also been demonstrated to benefit from optimal noise \cite{13}. Audhkhasi et. al. has shown that Convolutional Neural Networks reduces average per-iteration training-set cross entropy and the average per-iteration training-set classification error by 39\% and 47\%, respectively \cite{10}\cite{11}\cite{12}. Dropout is one of the most successful applications of noise in Deep Learning, and it involves randomly "dropping" a neuron from the network, which dramatically reduces overfitting \cite{14}. It can also be interpreted as injecting noise to hidden neurons during the ANN's training. Dropout is now a widely-used regularization technique deployed in many ANNs. We thus find that both biological and artificial neural networks benefit from carefully injected noise, which can actually serve to increase information integration in a physical system. 

\space
\bigskip

In contrast to a mounting body of evidence suggesting  that neuronal noise may play a significant role in cognitive development, categorical representation, and learning processes in the human brain, IIT penalizes a physical system for having noise, arguing that it would add entropy to the system, decrease information integration, and thus decrease $\phi$ \cite{17}. While this may conform to previous expectations of neuronal noise as a distraction at best, IIT's model does not agree with contemporary research on both artificial and biological neuronal noise, which demonstrates that it is not only vital for information integration, but plays a significant role in learning, development, and representation \cite{46}\cite{20}\cite{51}\cite{14}.

\space
\bigskip

Proponents of IIT will argue that neuronal noise is a detriment, not an asset: it increases entropy, which reduces information integration, as demonstrated by the success of the Neural Noise Hypothesis of Cognitive Decline. The Neural Noise hypothesis conjectures that noise in the brain increases with age, resulting in memory decline in aging patients\cite{19}. Voytek et. al. have demonstrated that $\frac{1}{f}$ electrophysiological noise increases with aging, supporting the neural noise hypothesis \cite{6}. Grossman et. al. have demonstrated that signals from a resting individual's brain predict their performance on visual recognition tasks, with higher amplitude fluctuations correlated with decreased performance. Grossman et. al. interprets these findings to suggest "the level of an individual's internal “noise” ... limits performance in externally oriented demanding tasks," which supports the neural noise hypothesis \cite{21}. Kali suggests that because Multiple Sclerosis, a disease which debilitates over 2.3 million individuals, involves deterioration of the myelin sheath, it may worsen with increased neuronal noise, thus supporting the Neural Noise Hypothesis \cite{7}. 

\space
\bigskip

With competing theories such as the Neural Noise theory of Cognitive Decline and the Bayesian Coding Hypothesis, the former which suggests that noise is at best a distraction and at worst an impediment, and the latter which suggests that noise is the fundamental means by which the brain effortlessly performs bayesian computations — we suggest that there is a Goldilock's balance to noise: too little can prevent learning or categorical representation and too much can result in cognitive decline or memory loss. Furthermore, both Artificial and Biological Neural Networks are shown to benefit from an optimal amount of noise — in the former, it reduces classification error and prevents overfitting; in the latter, it is critical to forming categorical representations and cognitive development. IIT cannot stand in its current form, for even though it acknowledges the neuronal noise hypothesis by penalized a physical system for increased noise, it does not account for recent research on stochastic facilitation (i.e., the benefits of noise in the human brain). IIT fails to take into account contemporary research demonstrating the utility of neuronal noise as a means of simplifying Bayesian computations, creating conscious precepts, learning categorical representations, and facilitating cognitive development. As such, IIT must be revised to adopt contemporary research on noise, which suggest noise is critical to forming conscious precepts. 

\section{Conclusion}
IIT is a valiant effort to create a full-fledged theory of consciousness. However, its proposed barometer of consciousness, $\phi$, is computationally impractical to calculate for most physical systems, and thus makes IIT’s clinical utility minimal. We have shown that IIT’s stance in regards to neuronal noise is inconsistent with experimental data demonstrating neuronal noise in the brain plays a critical role in learning, visual recognition, and even categorical representation. IIT predicts that entropy due to noise will reduce information produced by a physical system, which is inconsistent with Engel et. al. and Grossman et. al.’s data demonstrating that decision-related noise is a necessary condition for learning and visual recognition tasks. At best, IIT is a first attempt at a theory of consciousness, as it needs to be modified to accommodate new experimental evidence demonstrating the utility of noise in predicting task performance, facilitating cognitive development, and anticipating arousal state \cite{23}\cite{46}\cite{20}.

\end{document}